\documentclass[aps, prb, twocolumn, amsmath]{revtex4-2}
\usepackage [latin1]{inputenc}
\usepackage{amsmath}
\usepackage{graphicx}
\usepackage[dvipsnames]{xcolor}
\usepackage{dcolumn}
\usepackage{bm}
\usepackage[version=4]{mhchem}
\usepackage{makecell}
\usepackage[colorlinks=true, urlcolor=blue, linkcolor=black, citecolor=black]{hyperref}
\usepackage{xfrac}
\usepackage{mathtools}
\usepackage{comment}
\usepackage{tabularx}
\usepackage{physics}

\def\etal{\textit{et al}.}

\newcommand{\beq}{\begin{equation}}
\newcommand{\eeq}{\end{equation}}
\newcommand{\bea}{\begin{eqnarray}}
\newcommand{\eea}{\end{eqnarray}}

\begin{document}

\title{Dominant spin-spin relaxation mechanism at clock transition of the \ce{Ho_{x}Y_{1-x}W_{10}} complex at different concentrations}

\author{Xiao Chen,$^{1}$}
\author{Haechan Park$^{2}$}
\author{Silas Hoffman$^{2,3}$}
\author{Shuanglong Liu,$^{1,2}$}
\author{Hai-Ping Cheng$^{1,2}$}
\email{ha.cheng@northeastern.edu}

\affiliation{
$^1$Department of Physics, Northeastern University, Boston MA 02115 USA\\
$^2$Center for Molecular Magnetic Quantum Materials, University of Florida, Gainesville, Florida 32611, USA\\
$^3$Laboratory for Physical Sciences, 8050 Greenmead Dr., College Park, MD 20740}
\date{\today}

\begin{abstract}
   
Spin decoherence poses a significant challenge in molecular magnets, with the nuclear spin bath serving as a prominent source. Intriguingly, spin qubits at the clock transition exhibit remarkable insensitivity to the surrounding nuclear spins. Recent experimental studies have unveiled a correlation between the decoherence time and the density of spin qubits, prompting our investigation into the contribution of the qubit bath to spin decoherence. In this paper, we present a comprehensive theoretical analysis of a few S=1 spin qubits, focusing on their interaction at the clock transition. Employing the exact diagonalization and the cluster correlation expansion (CCE) method, we simulate the dynamics of spin decoherence while varying the density of the qubit bath. To ensure the realism of our simulations, we incorporate structural and energetic parameters derived from previous studies on the HoW10 crystal. Our findings indicate that when the energy mismatch between the energy splittings of two qubits exceeds their interaction strength, they can become effectively insensitive to each other, offering an explanation for the absence of observed changes in the T2 time during experiments with lower qubit densities. Understanding the role of qubit bath density in spin decoherence at the clock transition not only advances our knowledge of decoherence mechanisms but also provides insights for the development of strategies to protect coherence in molecular magnets and other quantum systems. By optimizing the density of spin qubits, we can enhance the coherence properties and pave the way for improved performance of quantum devices. Overall, this study offers valuable insights into the relationship between qubit bath density and spin decoherence at the clock transition, contributing to the broader understanding and control of quantum systems in molecular magnets.
\end{abstract} 

\maketitle 

\section{Motivation}
\begin{comment}
Molecular spins have been a promising candidate for applications in quantum information science by utilizing the electron spin as a computational bit\cite{eskridge2023nonperturbative,ardavan2006will,cruz2016influence,takahashi2011decoherence}. These bits are also called quantum bits or qubits in short. One of the biggest challenges using the qubits for quantum information processing is the phenomenon called decoherence \cite{2007,Stamp2008} where the quantum system loses its information to the environment due to the interaction with the surroundings. A major known source of decoherence is dipolar coupling to the nuclear spins. In order to protect the qubits, many solutions have been proposed\cite{Kaminski2014,Pla2012} but most of them rely on diluting the system to have less dipolar coupling between spins. Although the dilution of spins increase the coherence time, it's disadvantageous for building quantum computers as the dilution leads weaker interaction between qubits and they do need to interact to work as a multi-qubit processor. There's another route that has been proposed and studied where we don't need extreme dilution is to utilize systems with avoided crossing. When the system is at this avoided crossing, or also known as clock transition (CT), qubits are largely protected by local magnetic fluctuation induced by the nuclear spin bath\cite{Shiddiq2016,Kundu2023}.
\end{comment}
The field of quantum information science has seen a surge in interest, largely fueled by the promise of molecular spins. These molecular spins utilize the electron's intrinsic spin as the foundational building block for computation, commonly referred to as quantum bit, or qubits in short \cite{eskridge2023nonperturbative,ardavan2006will,cruz2016influence,takahashi2011decoherence}. However, a formidable challenge emerges - the perplexing phenomenon known as decoherence \cite{2007, Stamp2008}. Decoherence unfolds when the quantum system loses its information or quantumness to the surrounding environment through interactions. One of the primary sources responsible for this is the dipolar coupling with nuclear spins. While numerous strategies have been proposed to protect qubits against decoherence \cite{Kaminski2014, Pla2012}, many of them rely on a solution involving dilution, which effectively reduces the strength of dipolar coupling among spins. However, as coherence time is prolonged through spin dilution, an unsettling trade-off arises - weaker interactions between qubits. This trade-off poses a significant drawback for constructing robust quantum computers, as qubits need to interact effectively to function as a multi-qubit processor \cite{Shiddiq2016}.

However, there is an alternative approach available that bypasses the requirement for significant dilution. Instead, it centers on systems that exhibit avoided crossings of energy levels, commonly referred to as clock transitions (CT) because of their robust phase stability, inspired by the principles of atomic clocks. At these CTs, qubits find robust protection against decoherence caused by local magnetic fluctuations induced by the nuclear spin bath \cite{Shiddiq2016,Kundu2023}.

While extensive research has explored the impact of the nuclear spin bath, there remains a substantial gap in our knowledge. Surprisingly few studies have ventured into the effects of the electron spin bath at the CT. It is within this context that our study unfolds, as we embark on a comprehensive exploration of the various aspects of the electron spin bath's influence on the decoherence process at the CT. Notably, we deliberately exclude the nuclear spin bath from our considerations, choosing to focus solely on the contributions of the electron spin bath in unraveling the mysteries of decoherence in this specific quantum regime.

\section{Theory and Computational Details}

\subsection{Model Hamiltonian}
    In an effort to mimic the clock transition, we set up a model Hamiltonian containing $N$ number of $S=1$ electron spins with phenomenological longitudinal and transverse anisotropy.  The one spin Hamiltonian reads,
    \begin{equation}\label{eq:one_spin_ham}
        \hat{H}_{i}=D\left[\hat{S}_z^2-\frac{1}{3}S\left(S+1\right)\right]+E\left(\hat{S}_x^2-\hat{S}_y^2\right)+\gamma_e \left(B_0-B_{min}\right)\hat{S}_z,
    \end{equation}
    where $\gamma_e$ is the electron gyromagnetic ratio in units of MHz/T. $D$ and $E$ are the second-order zero-field splitting parameters whose values are -45 GHz and 4.5 GHz respectively for \ce{HoW10} \cite{Kundu2023} .$\hat{S}_i$ are $S=1$ spin operators along the $i-axis$. If we solve eqn. (\ref{eq:one_spin_ham}) at the clock transition with $B_0=B_{min}$, we get three following eigenstates,  $\ket{\uparrow}=\frac{1}{\sqrt{2}}\left(\ket{+1}+\ket{-1}\right)$, $ \ket{\downarrow}=\frac{1}{\sqrt{2}}\left(\ket{+1}-\ket{-1}\right)$, and $\ket{0}$ in the $\ket{m_z}$ basis. Their eigenvalues are $-\frac{1}{3}\abs{D}+E$, $-\frac{1}{3}\abs{D}-E$, and $+\frac{2}{3}\abs{D}$, respectively. Here, we choose the bottom two eigenstates as our qubit $\ket{\uparrow}$ and $\ket{\downarrow}$ states.
       
    For more than one spin, we add a dipolar interaction between spins.
    \begin{equation}
    \hat{H}=\sum_i^N \hat{H_i}+\sum_{i\neq j}\vec{S}_i\cdot\overleftrightarrow J_{ij}\cdot\vec{S}_j.
    \end{equation}
    For the dipolar interaction tensor, $\overleftrightarrow J$, we used a point dipole-dipole interaction that is in the following form.
    
    \begin{equation}
        \overleftrightarrow J=-\gamma_e\gamma_e \frac{\hbar^2}{4\pi\mu_0}\left[\frac{3\vec{r}_{ij}\otimes\vec{r}_{ij}-\abs{\vec{r}_{ij}}^2\boldsymbol{I}}{\abs{\vec{r}_{ij}}^5}\right]=\begin{pmatrix} J_{xx}&J_{xy}&J_{xz}\\J_{yx}&J_{yy}&J_{yz}\\J_{zx}&J_{zy}&J_{zz} \end{pmatrix}.
    \end{equation}
    The size of the Hamiltonian increases $3^N\times3^N$ with S=1 spins, which is disadvantageous for computation, we map our S=1 Hamiltonian onto the qubit basis. Using the superoperator defined as \cite{Kundu2023}
   \begin{equation}\label{eq:superoperator}
    \hat{\hat{P}}\hat{O}=\begin{pmatrix}
        \bra{\alpha}\hat{O}\ket{\alpha} &\bra{\alpha}\hat{O}\ket{\beta}\\
        \bra{\beta}\hat{O}\ket{\alpha}&\bra{\beta}\hat{O}\ket{\beta}
    \end{pmatrix}
    =\hat{o},
\end{equation}
we find the following projection rules \cite{Kundu2023}:
\bea
&&\hat{S}_z^2\rightarrow\boldsymbol{I}_{2\times2},\qquad \hat{S}_x^2-\hat{S}_y^2 \rightarrow \hat{\sigma}_z , \qquad\hat{S}_x \rightarrow 0,\qquad \hat{S}_y \rightarrow 0,\nonumber\\
&&\qquad \hat{S}_z \rightarrow \hat{\sigma}_x.
\eea
With these projection rules, we can rewrite the Hamiltonians. The projected hamiltonians read,
\begin{equation}\label{eq:projected_single_hamiltonian}
    \hat{h}_i=E\hat{\sigma}_z+\gamma_e(B_0-B_{min})\hat{\sigma}_x,
\end{equation}
\begin{equation}
       \hat{h}=\sum_i^N \hat{h}_i +\sum_{i\neq j} J_{ij;zz}\hat{\sigma}_x^i\hat{\sigma}_x^j.
\end{equation}
\begin{figure}[h]
    \centering
    \includegraphics[width=0.8\columnwidth]{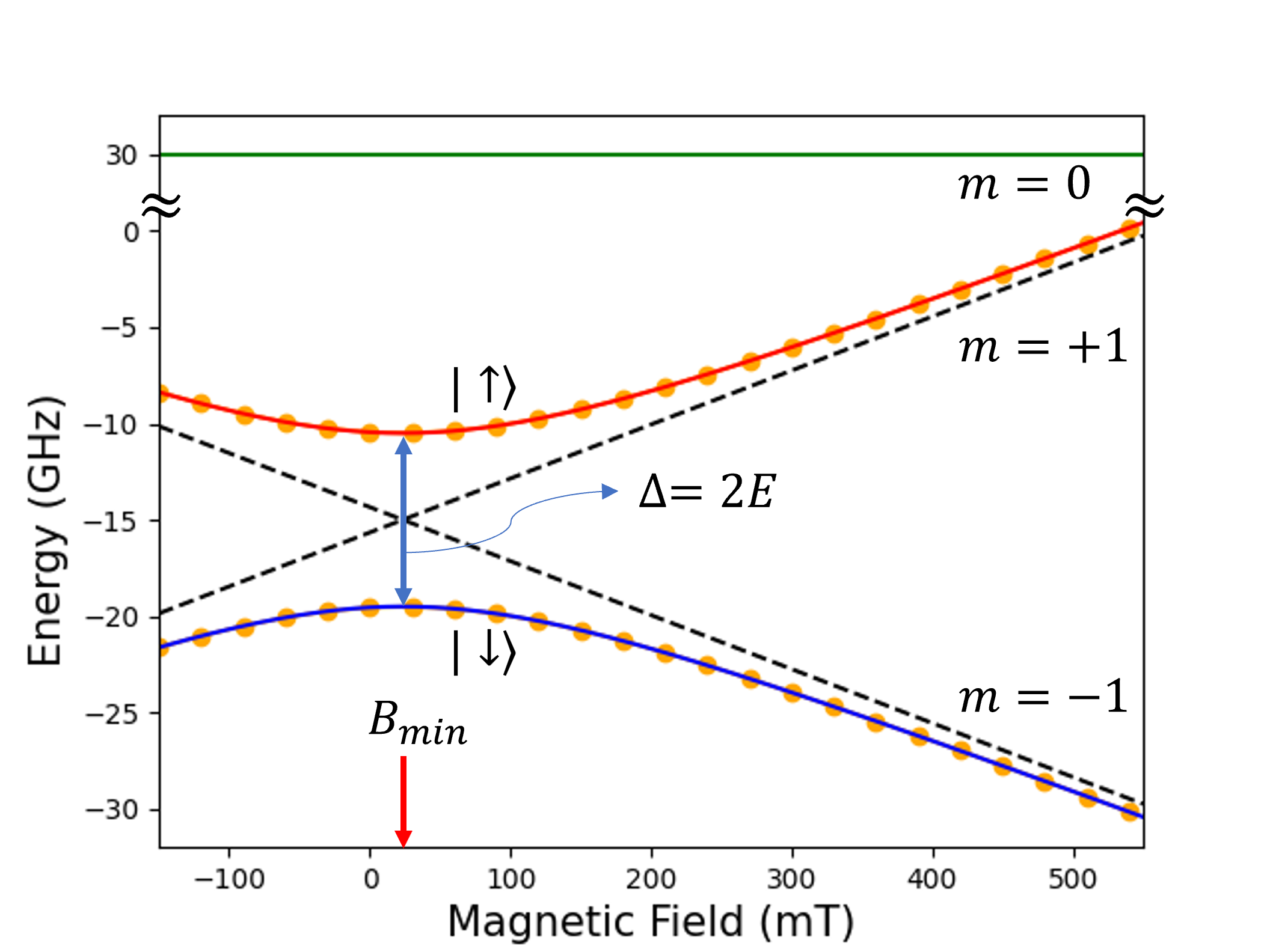}
    \caption{Zeeman diagram of the model Hamiltonian. Red and Blue curves indicate the qubit $\ket{\uparrow}$ and $\ket{\downarrow}$ states respectively. The clock transition is at $B_0=B_{min}=23.6mT$ and the qubit energy gap $\Delta=2E$. The dots represet the zeeman diagram of the projected Hamiltonian, which overlaps with the original Hamiltonian.}
    \label{fig:avoided_crossing}
\end{figure}
This new projected Hamiltonian, eq.\ref{eq:projected_single_hamiltonian}, produces the same Zeeman diagram up to a constant $(D/3)\boldsymbol{I}$ (See Figure \ref{fig:avoided_crossing}), which doesn't participate in the dynamics. In this study, we simulate the systems at CT so the magnetic field term in eq.(\ref{eq:projected_single_hamiltonian}) will not be used for the entire study.

\subsection{Spin Echo and Coherence Measurement}
    Pulsed electronic spin resonance (ESR) or electronic paramagnetic resonance (EPR) spectroscopy has been crucial for the measurement of coherence and implementation of spin-based qubit systems. On the Bloch sphere, the initial state is prepared to be along the external magnetic field direction, in this case, $z-axis$. The pulse sequence begins with a $\pi/2$-pulse which rotates this initial magnetization in an equilibrium direction into a transverse plane. This state can be represented by a superposition of the qubit $\ket{\uparrow}$ and $\ket{\downarrow}$ states.  After the $\pi/2$-pulse, the magnetization is allowed to precess around the $z-axis$ for the delay time, $\tau$. During this evolution, the magnetization can dephase due to noise sources, leading to a loss of the initial transverse magnetization. After the first delay, a $pi$-pulse is applied, which inverts the magnetization. In the ensemble picture, the most advanced spin in phase is now placed in the furthest behind all other spins and vice versa. After another delay time $\tau$ passes, the magnetizations refocus at the initial direction after the $\pi/2$-pulse where they started, followed by the measurement of the final magnetization, which provides us the information about the spin decoherence. All the pulses are assumed to be ideal pulses, where the pulse duration is infinitely short and the bandwidth is infinite. Since we project our Hamiltonian from \ref{eq:one_spin_ham} to \ref{eq:projected_single_hamiltonian}, it gives us a simple rotation framework with the Pauli matrices, $\hat{sigma}$, which conveniently rotates our qubit on the Bloch-sphere.
        A quantum system in a pure state $\ket{\Psi}$ can be described by a density operator, $\hat{\rho}=\ket{\Psi}\bra{\Psi}$. In the eigenstates $\ket{\uparrow}$ and $\ket{\downarrow}$, the density operator can be represented by a $2\times2$ matrix.
        \begin{equation}
            \hat{\rho}=\begin{pmatrix}
                \rho_{\uparrow\uparrow}&\rho_{\uparrow\downarrow}\\\rho_{\downarrow\uparrow}&\rho_{\downarrow\downarrow}
            \end{pmatrix}.
        \end{equation}
    The diagonal elements, $\rho_{\uparrow\uparrow}$ and $\rho_{\downarrow\downarrow}$, indicate the population of each eigenstate and the off-diagonal elements represent the coherence of the superposition state. In the qubit eigenstates, the state after the $\pi/2$-pulse that rotates spin by $pi/2$ around the $y-axis$ would be $\frac{1}{2}(\ket{\uparrow}+\ket{\downarrow})$. This means the magnetization is pointing $x-axis$. If we construct a density matrix of this state, it would be,
    \begin{equation}\label{eq:initial_density}
        \frac{1}{\sqrt{2}}\left(\ket{\uparrow}+\ket{\downarrow}\right)\frac{1}{\sqrt{2}}\left(\bra{\uparrow}+\bra{\downarrow}\right)=\frac{1}{2}\begin{pmatrix}
            1\\1
        \end{pmatrix}\begin{pmatrix}
            1&1
        \end{pmatrix}=\frac{1}{2}\begin{pmatrix}
            1&1\\1&1
        \end{pmatrix}
    \end{equation}
     which is the initial matrix for each electron in our simulations.  If we let $\ket{\psi_0^i}=\frac{1}{\sqrt{2}}\left(\ket{\uparrow}+\ket{\downarrow}\right)$ for the intitial state of the $i$th spin, the total initial state $\ket{\Psi_0}$ would be $\ket{\Psi_0}=\ket{\psi_0^1}\otimes\ket{\psi_0^2}\otimes\cdots\otimes\ket{\psi_0^N}$ for $N$ spins. Then the total initial density is $\hat{\rho}_0=\ket{\Psi_0}\bra{\Psi_0}$. Once we prepared our initial density, we let the system evolve freely, using the time evolution operator for the second step of the Hahn-echo sequence \cite{hahn}. At $t=\tau$, the total density operator reads,
     \begin{equation}
         \hat{\rho}(\tau)=\hat{U}(\tau)\hat{\rho}_0\hat{U}^{\dag}(\tau),
     \end{equation}
     where $\hat{U}(\tau)=\exp{-i\hat{h}\tau}$ with $\hat{h}$ being the total Hamiltonian. As described above, the ideal $\pi$-pulse would then be applied to all the spins. The density operator reads
     \begin{equation}
         \hat{\rho}(\tau+)={\hat{P}_\pi^x}\hat{\rho}(\tau){\hat{P}_\pi^{x\dagger}}.
     \end{equation}
     Here  $+$ sign indicates the time right after the pulse is applied.
    Another time evolution takes place as the last sequence and the final density operator would be
    \begin{equation}
        \hat{\rho}(2\tau)=\hat{U}(\tau)\hat{\rho}(\tau+)\hat{U}^{\dag}(\tau).
    \end{equation}
    At the refocus step of the Hahn-Echo sequence, we measure the expectation value of $\hat{S_x}$ operator of the spin that we are measuring the coherence of. Therefore coherence can be in the form,
    \begin{equation}\label{eq:coherence_df}
        L(2\tau)=\frac{\left<\hat{S}_x(2\tau)\right>}{\left<\hat{S}_x(0)\right>}=\frac{tr\left[\hat{\rho}(2\tau)\hat{S}_x\right]}{tr\left[\hat{\rho}(0)\hat{S}_x\right]}
    \end{equation}
    We can do this procedure by making the $\hat{\rho}(2\tau)$ the reduced density matrix for the central spin at the last step.
    At $t=0$, we can quickly calculate and see $L(0)=1$, using eq. (\ref{eq:initial_density}) and eq. (\ref{eq:coherence_df}). 
When the coherence of the central spin decays, the rate of this decay is characterized by a time constant $T_2$ and will be fitted by the stretched exponential in the form of
\begin{equation}
    L(2\tau)=\exp{-\left(\frac{2\tau}{T_2}\right)^\beta},
\end{equation}
where $\beta$ is a constant parameter for the fitting. 
\subsubsection{Structural Information}
    Our study aims to have a model study but for more realistic and physical simulations, we adopt the \ce{Ho10W} crystal structure. We first took \ce{ErW10} structure from the work by AlDamen \etal \cite{AlDamen2008} and replaced one of the Er atoms in the unit cell with a Homium atom and the other with a Yttrium atom as the experiments were done with different concentration of Ho in the \ce{Ho_x Y_{(1-x)}W10} crystal. Then the structure was relaxed using first-principles calculation at the level of density functional theory(DFT) implemented in Vienna \textit{Ab Initio} Simulation Package (VASP) \cite{Kresse1993,Kresse1999}. We didn't optimize the cell parameters so they are the original values. The cell vectors are $\vec{a}=(12.74,\: 0.00,\: 0.00)$, $\vec{b}=(0.27,\: 13.07,\: 0.00)$, and $\vec{c}=(5.46,\: 2.42,\: 19.58)$ in Cartesian basis. Upon relaxation, we remove all the atoms except the Ho and Er atoms in the cell. 
  \begin{figure}[ht]
    \centering
    \includegraphics[width=0.8\columnwidth]{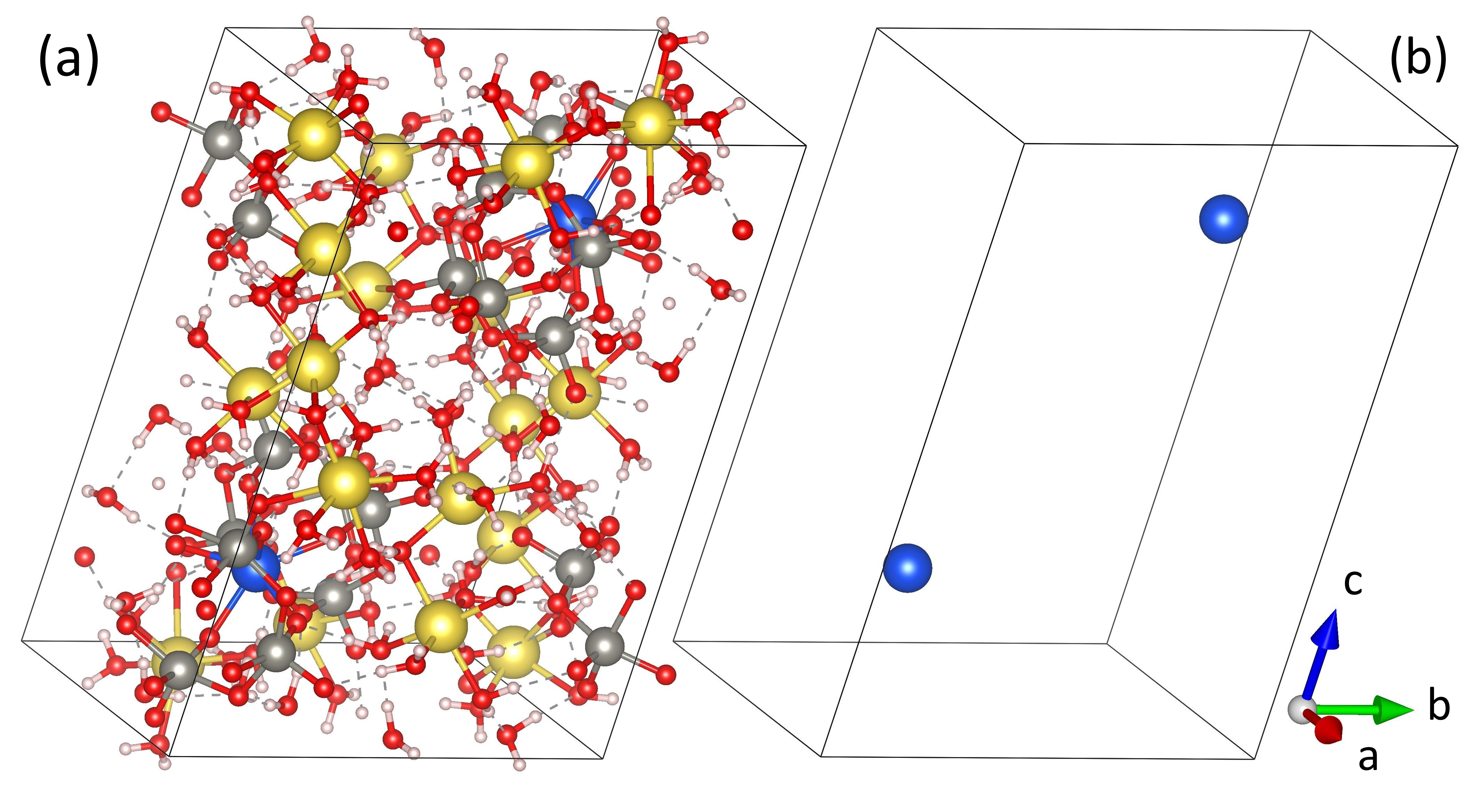}
    \caption{(a) the unit-cell of crystal \ce{HoW10} structure with all the atoms.(b) the same unit-cell with only the \ce{Ho} sites. This cell will be repeated to build supercells. The two two host positions are filled with Ho atoms. }
    \label{fig:how10structure}
\end{figure}

\begin{figure}[ht]
    \centering
    \includegraphics[width=0.5\columnwidth]{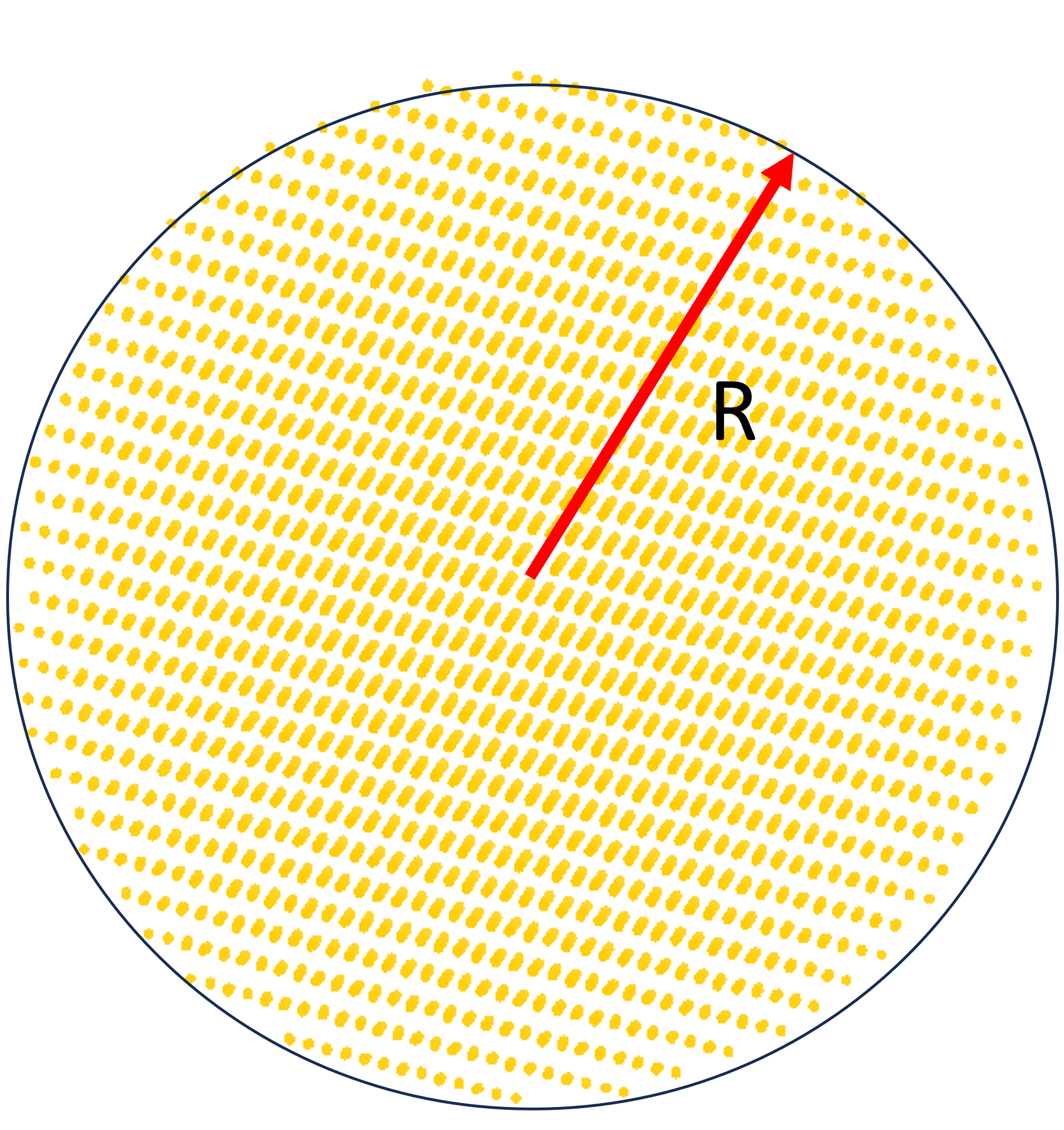}
    \caption{Schematics of the spherical box with radius of $R$. $R$ is determined by the number of spins we want to generate with respect to the density $x$.}
    \label{fig:spherical}
\end{figure}

    Using this Holumium and Erbium positions in the unit cell as shown in Figure \ref{fig:how10structure}, we build different sizes and shapes of supercells. We first try a spherical cell (See Figure \ref{fig:spherical} starting with $x=0.001$ concentration, or density. If we want to use 6 spins to do our simulationn, this would mean we need a supercell that contains 6000 empty sites to populate our 6 spins to meet $x=0.001$ density. When increasing the density, the natural way to do is to keep the cell and put more spins in the cell but this is computationally infeasible without any approximations, such as Cluster Correlation Expansion (CCE) \cite{Yang2008,Onizhuk2021}. Alternatively, we shrink the cell size to increase the density. For example, for density $x=0.01$, instead of populating 60 spins in the 6000 sites, we still generate 6 spins out of 600 sites. The size of the cell will also be smaller according to the density. Our central spin will always be located at the center of the sphere. 

\begin{table}[htbp]
  \caption{Radius $R$ and number of sites in the spherical box as a function of density.}
  \label{radius}
  \begin{ruledtabular}
  \begin{tabular}{lrr}
    Density $x$ & $R$ (\AA) & Number of sites \\
    \hline
    0.001 & 132.66 & 6002 \\
    0.002 & 105.30 & 3000 \\
    0.003 &  92.20 & 2001 \\
    0.004 &  83.30 & 1499 \\
    0.005 &  77.70 & 1201 \\
    0.006 &  72.78 & 1000 \\
    0.007 &  69.63 &  858 \\
    0.008 &  66.12 &  750 \\
    0.009 &  63.91 &  668 \\
    0.010 &  61.64 &  601 \\
    0.100 &  29.00 &   61 \\
  \end{tabular}
  \end{ruledtabular}
\end{table}

\begin{figure}[ht]
    \centering
    \includegraphics[width=\columnwidth]{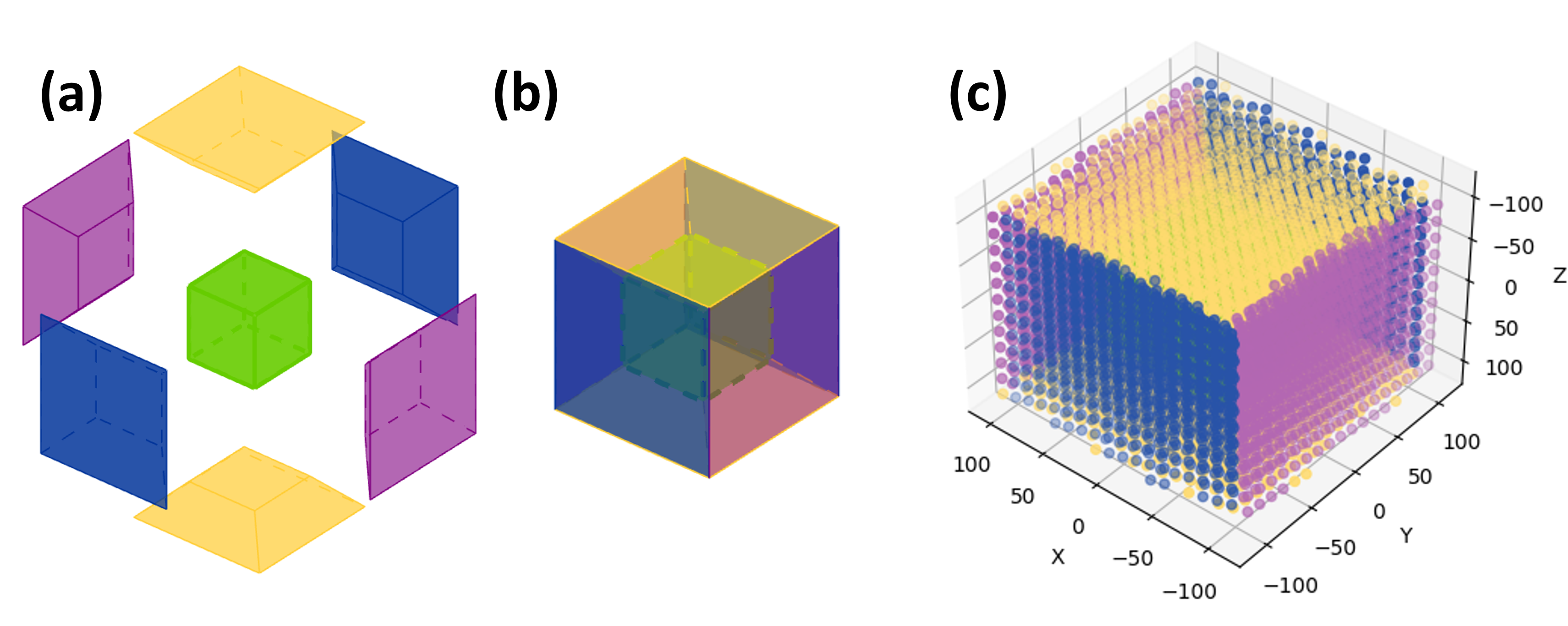}
    \caption{(a)Schematics of the dissected square when separated. (b) Dissected square when assembled. (c)  Actual square cell that is color-coded to match the schematics. }
    \label{fig:squarecell}
\end{figure}

To study the effect of spatial uniformity of spin configurations later, we build a dissected cube as shown in Figure \ref{fig:squarecell}. Each section in the cube will have almost the same volume/number of sites. At the center of the center block we have our central spin and no other spins will be populated in the central section. When using this cell, 6 out of a total of 7 spins will be randomly populated in each of 6 sections as the central spin position is fixed at the center.

\subsection{Cluster Correlation Expansion}
Here we provide a brief overview of cluster correlation expansion (CCE) technique. The CCE was originally introduced in reference \cite{Yang2008}. The fundamental concept behind the CCE approach is that the decoherence induced by the spin bath can be broken down into a series of irreducible contributions from clusters of bath spins. When expressed in terms of the coherence function:

\begin{equation}\label{eq:clusterexpansion}
    L(\tau)=\prod_C \tilde{L}(\tau)=\prod_i \tilde{L}_{\left\{i\right\}}\prod_{i,j}\tilde{L}_{\left\{i,j\right\}}\cdots
\end{equation}
Each cluster contribution is recursively defined as:
\begin{equation}
    \tilde{L}=\frac{L_C}{\prod_{C'\subset C}L_{C'}}
\end{equation}

where $L_C$ represents the coherence function of the qubit, interacting solely with the bath spins within a specific cluster $C$ that is characterized by the cluster Hamiltonian that governs the dynamics of the cluster $C$, and $\tilde{L}_C'$ corresponds to the contributions of the subcluster $C'$ within $C$. For instance, the contribution of a single spin $i$ is equivalent to the coherence function of the bath with a solitary isolated spin $i$:
\begin{equation}
    \tilde{L}_{\left\{i\right\}}=L_{\left\{i\right\}}.
\end{equation}
The contribution of a pair of spins $i$ and $j$ is given by:
\begin{equation}
    \tilde{L}_{\left\{i,j\right\}}=\frac{L_{\left\{i,j\right\}}}{L_{\left\{i\right\}}L_{\left\{j\right\}}},
\end{equation}
and so forth.

The maximum cluster size included in the expansion determines the order of the CCE approximation. For instance, in the CCE2 approximation, contributions are considered up to spin pairs, while in CCE3, triplets of bath spins are taken into account, and so on. 

\section{Results and Discussion}
\subsection{Two-Spin Analysis}
\begin{figure}[ht]
    \centering
    \includegraphics[width=\columnwidth]{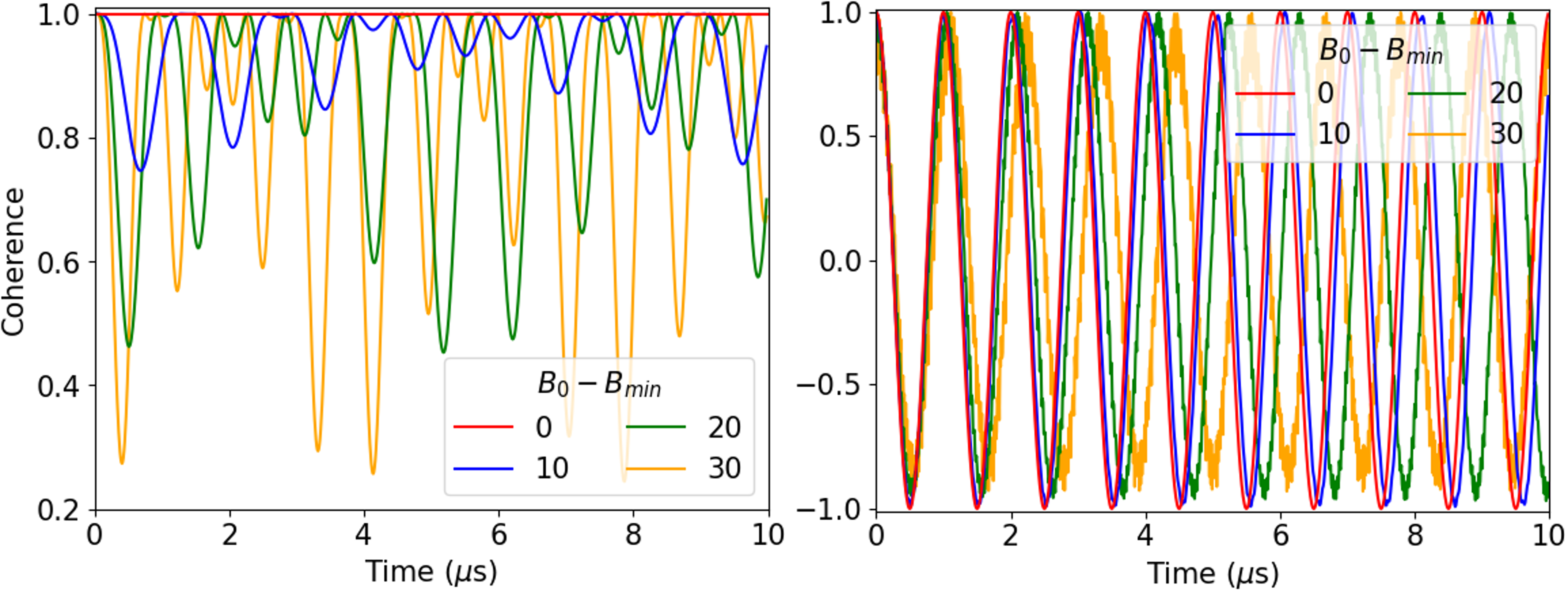}
    \caption{(a) coherence signals between one electron spin and one proton spin in different magnetic fields $B-B_{min}$ in units of mT. As the field approaches to CT, the signal amplitude decreases and disappears at CT, which makes the proton bath invisible to the electron spin. (b) coherence signals between two electron spins near CT magnetic fields. In this case, the signal amplitude decreases as the field moves away from CT. Interaction between the two electron spins is set to 1 MHz which matches the frequency of the signal cosine function.}
    \label{fig:protonvselectron}
\end{figure}

In this section, we are going to explore the simplest case with two electron spins. Before we begin the discussion of the two-electron spin analysis, a brief comparison will be given between the two-electron spin case and one electron and one proton case near CT. As shown in Figure \ref{fig:protonvselectron} (a), the coherence signal amplitude between the electron spin and the proton spin decreases as the system gets closer to CT and the proton spin becomes completely invisible to the electron spin. On the other hand, the coherent signal amplitude doesn't change much near and at CT. In fact, the amplitude gets smaller as it goes away from CT. These results tell us that even at CT, the interaction between electron spins can have an effect on the overall coherence curve.  
We begin our discussion with the simplest case of one central spin and one bath spin. We call spins that are not the central spin bath spins because their coherences are not measured, althouh they are all the same species of electron spins. All the bath spins will also feel the same pulse as the central spin. 
  
%두개 스핀 케이스 넣어야할 것, b=0아닌 케이스에서 b=0으로 갈때 차이점, 그리고 E가 같을 때부터 같지 않을때의 차이를 그림 그렿 놓은것

To understand this simple result, let us write down the 2 spin Hamiltonian explicitly:
\begin{equation} \label{eq:2spin_ham}
    \hat{H}_{2psin}=2E(\hat{S}_{z0}+\hat{S}_{0z})+4J\hat{S}_{xx},
\end{equation}
 where $\hat{S}_{ij}=\hat{S}_i\otimes\hat{S}_j$ with $i=0,x,y,z$ represent $2\times2$ Spin-1/2 operators and a $2\times2$ for $i=x,y,z$ and identity matrix for $i=0$. Note here that there are additional factors of 2 and 4 before $E$ and $J$ respectively. This is because $\hat{S}_{i=x,y,z}=\frac{1}{2}\hat{\sigma}_{i=x,y,z}$.% and when $i\neq0$ and $j\neq0$, we have a factor of 2 for the notation. For example, $\hat{S}_{xx}=2\hat{S}_x\otimes\hat{S}_x$% and this is why the $\hat{S}_{xx}$ term has a factor of $2$ instead of $4$ in eq. (\ref{eq:2spin_ham}).
The initial density operators before the $\pi/2$-pulse for each spin A and B are
\begin{equation}
    \ket{\Psi}_0^{A,B}=\begin{pmatrix}
        1\\0
    \end{pmatrix},\hat{\sigma}^{A,B}_0=\ket{\Psi_0^{A,B}}\bra{\Psi_0^{A,B}}=\begin{pmatrix}
        1&0\\0&0
    \end{pmatrix}=\hat{S}_z+\frac{1}{2}\hat{S}_0.
\end{equation}
We can see that the density operator can be represented with spin operators. The total initial density would be

\bea
        \hat{\sigma}(0)&=&\hat{\sigma}_0^A\otimes\hat{\sigma}_0^B=\left(\hat{S}_z+\frac{1}{2}\hat{S}_0\right)\otimes\left(\hat{S}_z+\frac{1}{2}\hat{S}_0\right)\nonumber\\
        &=&\frac{1}{2}\left(2\hat{S}_{zz}+\hat{S}_{z0}+\hat{S}_{0z}+\frac{1}{2}\hat{S}_{00}\right)
\eea
After the $\pi/2$-pulse, the density is prepared to be
\bea\label{eq:initial_density}
    \hat{\sigma}(0+)&=&e^{-i\frac{\pi}{2}\left(\hat{S}_{y0}+\hat{S}_{0y}\right)}\hat{\sigma}(0)e^{i\frac{\pi}{2}\left(\hat{S}_{y0}+\hat{S}_{0y}\right)}\nonumber\\
    &=&\frac{1}{2}\left(2\hat{S}_{xx}+\hat{S}_{x0}+\hat{S}_{0x}+\frac{1}{2}\hat{S}_{00}\right)
\eea
Our numerical and analytical result shows $\hat{S}_{xx}$ doesn't participate in the dynamics. $\hat{S}_0$ term also doesn't affect the coherence as it's an identity matrix. 
The time evolution operator can be written \cite{Mathematica}

\[    \hat{U}(\tau)=e^{-i\tau\hat{H}_{2spin}}\]
\begin{equation}
\resizebox{\columnwidth}{!}{$
\left(
\begin{array}{cccc}
 \cos \left(\frac{\tau \sqrt{\xi}}{2}\right)-\frac{4 i E \sin \left(\frac{\tau \sqrt{\xi}}{2}\right)}{\sqrt{\xi}} & 0 & 0 & -\frac{i J \sin \left(\frac{\tau \sqrt{\xi}}{2}\right)}{\sqrt{\xi}} \\
 0 & \cos \left(\frac{J \tau}{2}\right) & -i \sin \left(\frac{J \tau}{2}\right) & 0 \\
 0 & -i \sin \left(\frac{J \tau}{2}\right) & \cos \left(\frac{J \tau}{2}\right) & 0 \\
 -\frac{i J \sin \left(\frac{\tau \sqrt{\xi}}{2}\right)}{\sqrt{\xi}} & 0 & 0 & \cos \left(\frac{\tau \sqrt{\xi}}{2}\right)+\frac{4 i E \sin \left(\frac{\tau \sqrt{\xi}}{2}\right)}{\sqrt{\xi}} \\
\end{array}
\right)
$}
\label{eq:time_evolution}
\end{equation}

 where $\xi=4E^2+J^2$.
Once the system is prepared at $t=0+$, we evolve the system for $\tau$ and the density operator will evolve with the system hamiltonian. 
\begin{equation}
    \hat{\sigma}(\tau)=\hat{U}(\tau)\hat{\sigma}(0+)\hat{U}(\tau)^\dag.
\end{equation}
%The full expression can be found in the appendix eq. (\ref{eq:dmtau}).
According to the Hanh-Echo sequence, the $\pi$-pulse, which is defined to be,

\begin{equation}
   \hat{P}_\pi^x=\exp{-i\pi\left(\hat{S}_{x0}+\hat{S}_{0x}\right)}
  =\left(
\begin{array}{cccc} 0& 0& 0&-1 \\ 0& 0&-1& 0 \\ 0&-1& 0& 0 \\-1& 0& 0& 0 \\\end{array} \right),
\end{equation}
is applied and  the density operator after the $\pi$-pulse is denoted by 
$\hat{\sigma}(\tau+)=\hat{P}^{x}_\pi\hat{\sigma}(\tau){\hat{P}_\pi^{x\dag}}$.
Then the sequence is completed by another time evolution by $\tau$. %The density operator at $t=2\tau$ can be also found in the appending eq.(\ref{eq:dm2tau}).
Finally, we measure the coherence using eq.(\ref{eq:coherence_df}) and get,
\bea\label{eq:l2tau}
L(2\tau)&=&\frac{4 E^2 \cos ( J2\tau)}{4 E^2+J^2}+\frac{J \sin ( J2 \tau) \sin \left(2 \tau \sqrt{4 E^2+J^2}\right)}{\sqrt{4 E^2+J^2}}\nonumber\\
&&+\frac{J^2 \cos (2 J \tau) \cos \left(2 \tau \sqrt{4 E^2+J^2}\right)}{4 E^2+J^2}.
\eea
Since $E>>J$, we can approximate $L(2\tau)=\cos\left(J2\tau\right)$ and this analytically shows our simple result. 
Another result we want to present in the two-spin case is a situation where the two spins have slightly different qubit gaps. For this, we use the following Hamiltonian.
\begin{equation}\label{eq:gap_ham}
    \hat{H}_{gap}=2\left(E+\frac{\delta}{2}\right)\hat{S}_{z0}+2\left(E-\frac{\delta}{2}\right)\hat{S}_{0z}+4J\hat{S}_{xx}.
\end{equation}
$d$ is the difference in the qubit gaps between the two spins. This represents the case of two off-resonant spins. We observe that the signal amplitude becomes smaller as they are more off-resonant. The frequency also shifts from the perfectly resonant case. 
 \begin{figure}[ht]
    \centering
    \includegraphics[width=\columnwidth]{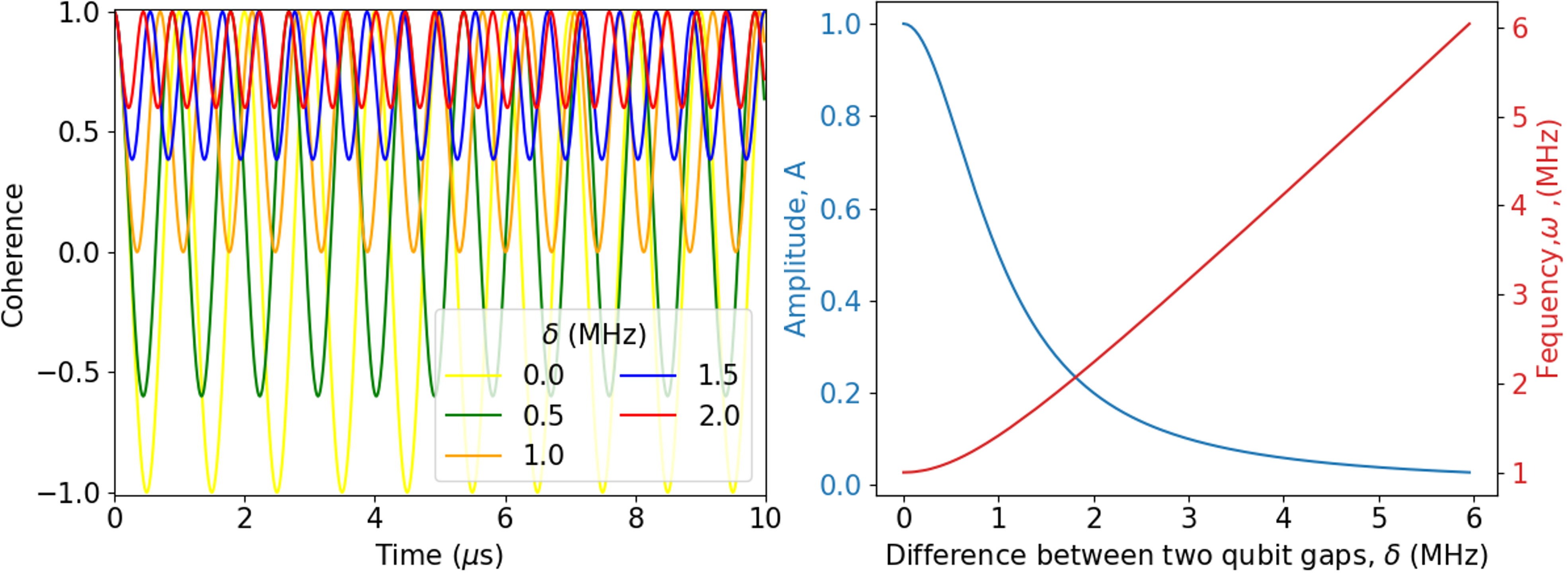}
        \caption{Left: Coherence signals with different qubit gaps, $\delta$ in units of MHz. When $d>J=1$MHz, all the signal values are positive. Right: The Amplitudes and frequencies as a function of the gap, $\delta$.}
    \label{fig:off-resonance-two-spin}
\end{figure}
If we define our signal with a perfect cosine function with a scalar shift, $B$, we can write down the signal function as follows.
\begin{equation}
    L(2\tau)=A\cos\left(2\omega\tau\right)+B.
\end{equation}
Changes of the coherence signals are plotted in Figure \ref{fig:off-resonance-two-spin}. The data points are perfectly fitted by the following analytical expressions so the fittings are not plotted. The coherence signal according to the Hamiltonian in eq.\ref{eq:gap_ham} is,
\bea
    L(2\tau)&=&\frac{4 \delta^2 E^2}{4 \delta^2 E^2+\delta^2 J^2+4 E^2 J^2+J^4}\nonumber\\
    &&+\frac{4 E^2 J^2 \cos \left(2 \tau  \sqrt{\delta^2+J^2}\right)}{4 \delta^2 E^2+\delta^2 J^2+4 E^2 J^2+J^4}+\cdots.
\eea
The smaller contributions are ignored here. We can immediately see that when $\delta=0$, this goes back to the eq.\ref{eq:l2tau} as expected. In the week coupling limit ($E>>J$), this can be further simplifed as,
\begin{equation}
    L(2\tau)\approx\frac{\delta^2}{\delta^2+J^2}+\frac{J^2}{\delta^2+J^2}\cos\left(2\tau\sqrt{\delta^2+J^2}\right).
\end{equation}
The amplitude of the signal decreases when two spins are not on perfect resonant. Since at $t=0$, the coherence starts at 1 this means the coherence values will be all positive due to the shrinking signal amplitude. This happens when the qubit gap difference, $\delta$ is greater than the interaction strength $J$. This fact will be used in the discussions later. The amplitude falls at a rate of $J^2/(\delta^2+J^2)$. The signal frequency also shifts with the $d$ following $\sqrt{\delta^2+J^2}$.

\subsection{N-Spin Simulation Results}
To study N-spins, the first crucial step involves determining the appropriate number of spins, denoted as N, required to yield meaningful results. This decision is of importance given that real-world systems can contain an enormous quantity of spins. Our approach begins by creating a spherical cell containing 3000 sites. At a density of x=0.001, this requires 3 spins in the cell. Subsequently, we expand the cell's radius until it accommodates 4000 sites, at which point we need a total of 4 spins.

Instead of randomly distributing all 4 spins across the entire cell, our method involves retaining the positions obtained from the initial 3-spin simulations (which are generated randomly) and designating one site within the newly added region of the cell for the placement of a spin. This newly incorporated area constitutes the entire cell minus the 3-spin cell, containing approximately 1000 sites. We proceed with this process incrementally, gradually increasing the spin count until we reach 10 spins. Clearly, the coherence will be influenced more by the specific positions of each spin rather than factors such as the number of spins within a given configuration. Therefore, we conduct Monte Carlo simulations in which we generate multiple random configurations. This enables us to observe the impact of varying the number of spins on the outcome. This is also similar to what is measured in the experiment. All electron spins will have very different configurations of other spins and they produce the echo signals. By averaging the signals, we can take this into account.
 \begin{figure}[ht]
    \centering
    \includegraphics[width=\columnwidth]{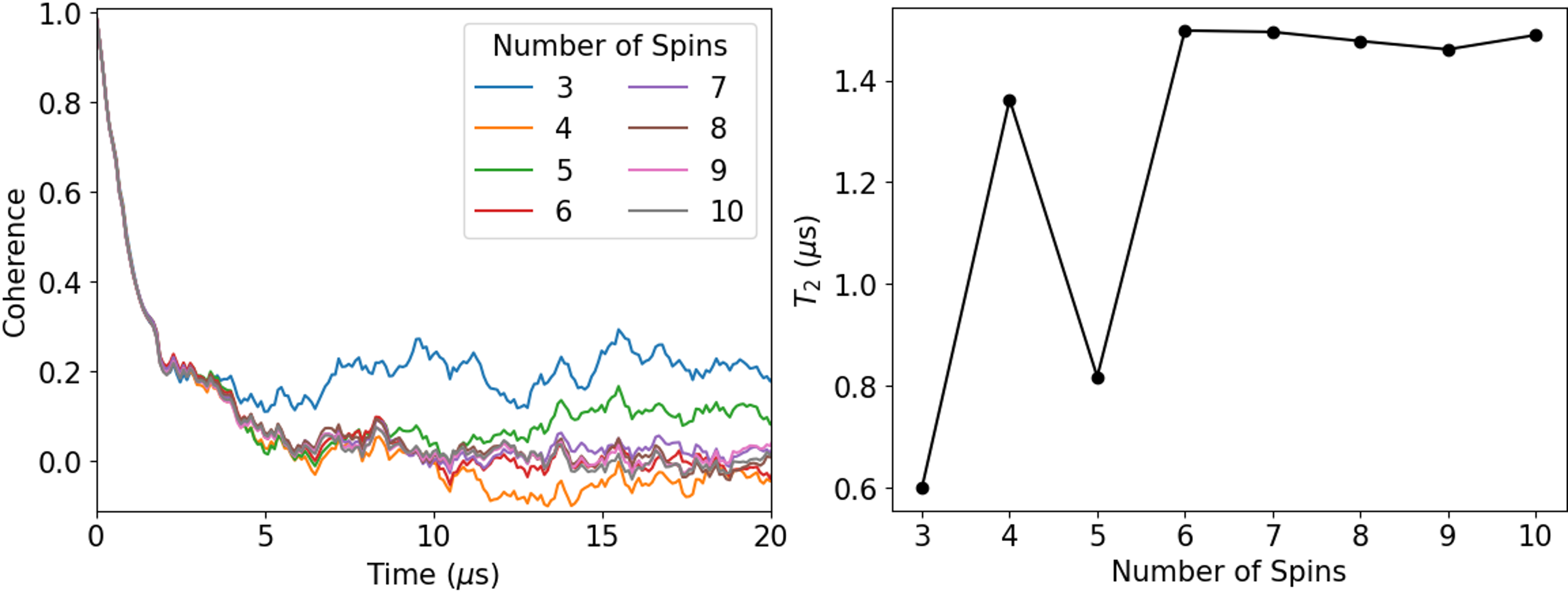}
    \caption{Left:The coherence signals of different number of total spins. Right: Their $T_2$ times fitted with a stretched exponential function.}
    \label{fig:num_spin_conv}
\end{figure}
In Figure \ref{fig:num_spin_conv}, we see that 3,4, and, 5 spins are deviated from 6 through 10 spin results. This also can be confirmed by the right figure where $T_2$ time is plotted as a function of the total number of spins. From 6 spins, there's almost no significant difference in $T_2$ time. Therefore we choose 6 spins for computational efficiency. All the $T_2$ curves are fitted to a stretched exponential  function defined as,
\begin{equation}
    f(2\tau)=\exp{-\left(\frac{2\tau}{T_2}\right)^\beta},
\end{equation}
 where $\beta$ is a constant parameter to be fitted.
We also need to verify the number of calculations required to stop the Monte Carlo simulations. Observing that above approximately 200 calculations, all the $T_2$ times exhibit a relatively stable convergence. 
 \begin{figure}[ht]
    \centering
    \includegraphics[width=0.6\columnwidth]{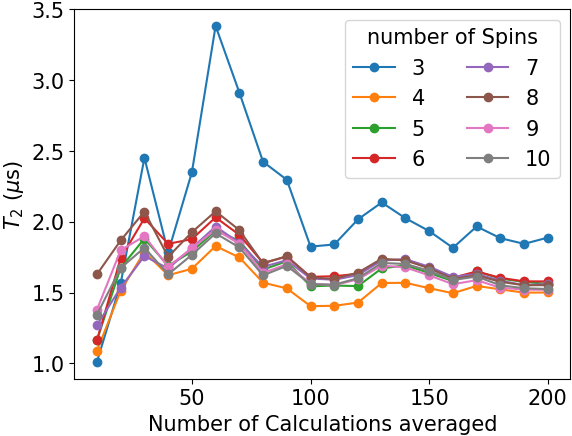}
    \caption{$T_2$ times are obtained from the averaged signals.}
    \label{fig:ncal}
\end{figure}

\begin{figure}[ht]
    \centering
    \includegraphics[width=\columnwidth]{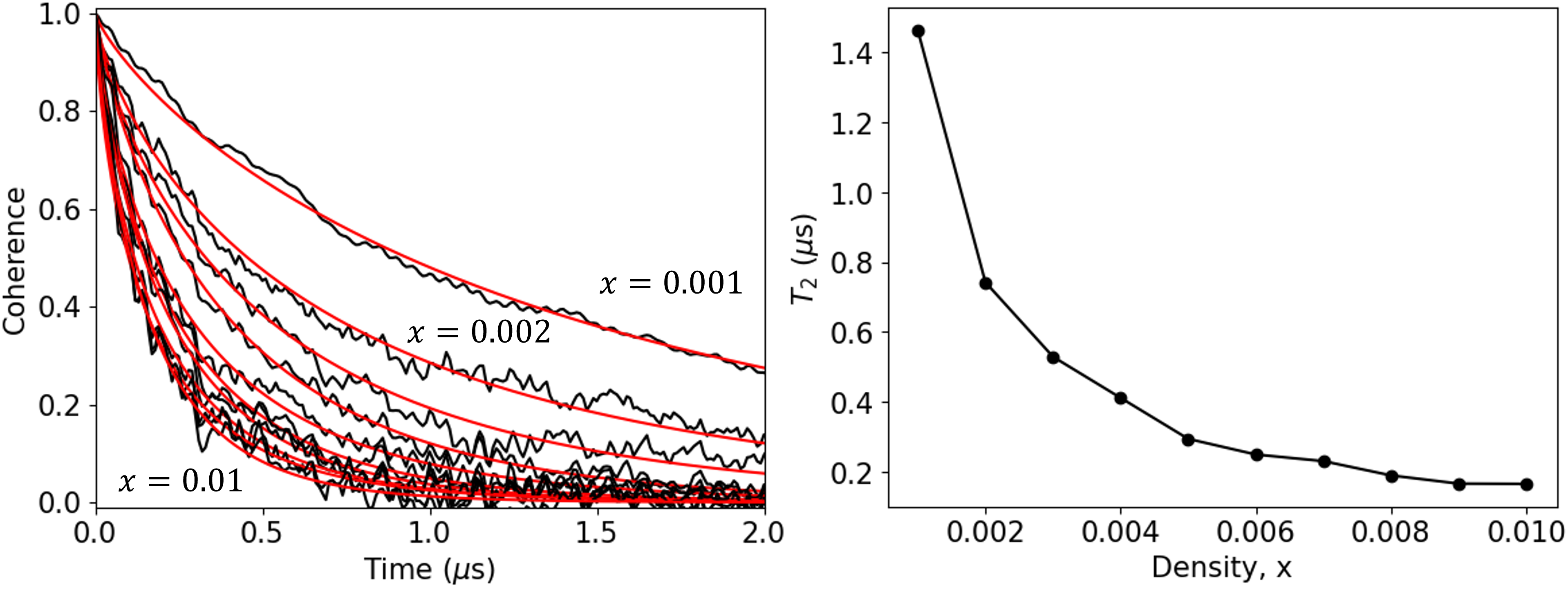}
    \caption{Left: coherence signals with different density of spin bath. The actual data are in black and the fitted curves in red. Right: $T_2$ time as a function of the density, $x$.}
    \label{fig:density}
\end{figure}

We are now prepared to explore the density effect. As previously mentioned, our approach to achieving higher density involves reducing the cell size rather than increasing the number of spins within the cell. The radii of the cells corresponding to densities are tabulated in Table. \ref{radius}. As illustrated in Figure \ref{fig:density}, the $T_2$ relaxation time exhibits an exponential decline with increasing density. This phenomenon aligns with our expectations, given that as the spins approach closer to one another, their interactions intensify, resulting in a more rapid decay of coherence.
However, as per Kundu et al., the experimental $T_2$ time at $x=0.001 $ is 8.4 $\mu s$, which is significantly slower than what our simulation results indicate. As per Kundu et al., the experimental $T_2$ time at x=0.001 is approximately 8 $\mu s$, which significantly differs from our simulation results. This discrepancy may arise from various sources, including the actual interaction strength compared to the point dipole-dipole approximation, which neglects the influence of other nuclear spins in the cell. Consequently, we embarked on an investigation focusing on scenarios where bath spins exhibit slight variations in their qubit gaps relative to each other.

\begin{figure}[ht]
    \centering
    \includegraphics[width=\columnwidth]{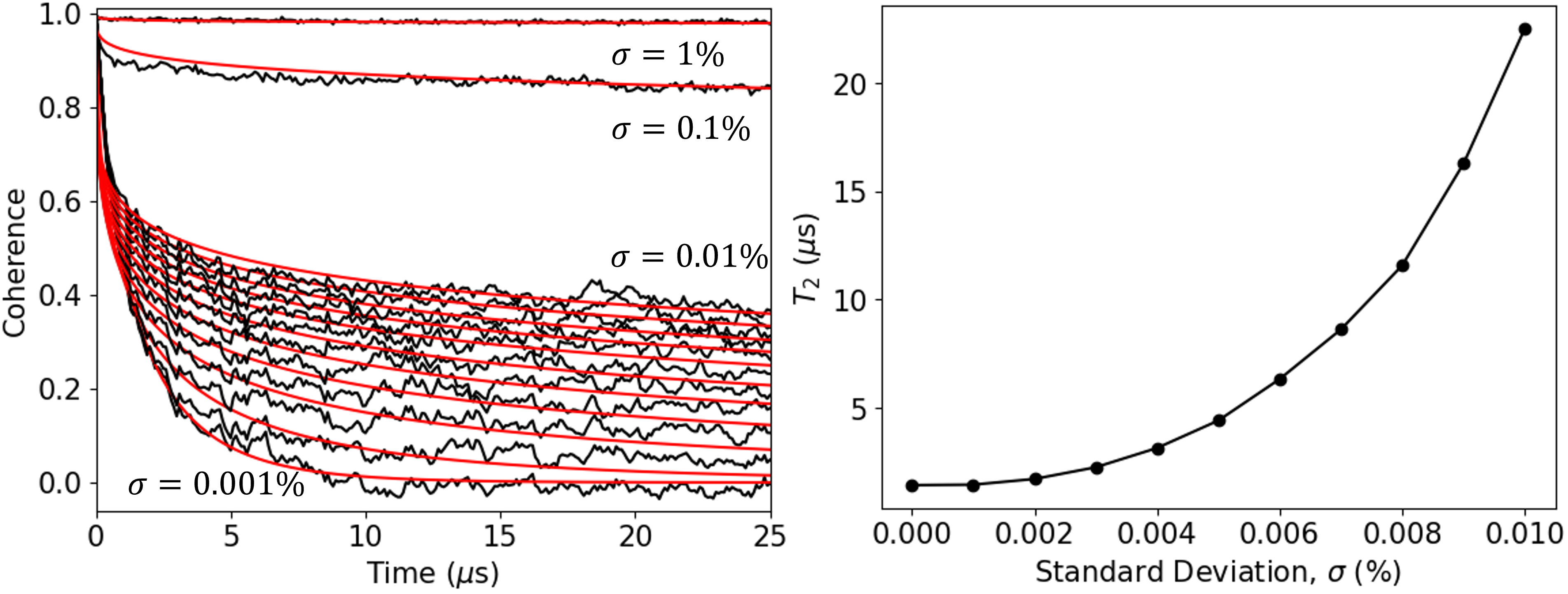}
    \caption{Left: coherence signals with different standard deviations of normal distributions of random qubit gaps. Again, the actual data are in black and the fitted curves in red. Right: $T_2$ time as a function of the standard deviation as a fraction of the original $E$.}
    \label{fig:off-resonance}
\end{figure}

For this simulation, we maintain the qubit gap of the central spin, denoted as E, at a constant value of 4.5GHz while assigning random numbers to the other spins. These random numbers follow a normal distribution centered around E=4.5 GHz, with varying standard deviations. The normal distribution used has the form,
\begin{equation}
    f\left(x\right)=\frac{1}{E\sigma \sqrt{2\pi}}\exp{-\left(\frac{\left(x-E\right)}{E\sigma}\right)^2}.
\end{equation}
The results are depicted in Figure \ref{fig:off-resonance}. What we observe is that the coherence decays at a slower rate when the spins are further off-resonance from each other. Our interpretation of this phenomenon is that as the spins become more off-resonant, the occurrences of flip-flops between spins become less frequent. In the extreme scenario where the energy gaps between spins are significantly disparate, the signal remains close to its initial value, effectively making the spins invisible to each other. This was evident in the case of interaction between an electron spin and a proton nuclear spin at CT. Numerically, when two spins possess distinct qubit gaps, we have already noted that the signal amplitude diminishes. Consequently, this results in smaller negative signals, and when these calculations are averaged, it leads to higher values, thus resulting in a longer decoherence time. When spins do not share an identical gap, the signal curves exhibit a consistent pattern characterized by a rapid initial decay followed by a slower, more gradual decline. In our subsequent simulations conducted over longer time intervals, the signals continue to exhibit a decay rather than converging to constant values.

Furthermore, experimental data reveals that there is no significant change in the $T_2$ time when transitioning from x=0.001 to x=0.01. Our findings allow us to propose an explanation for the disparity between our simulations and the empirical results. Small variations in the coordination environment and other local factors can introduce some degree of inhomogeneity in the qubit gaps. In cases where there is randomness in these qubit gaps, our simulations indicate that this can actually result in an increase in the $T_2$ time. This observed discrepancy might account for the substantial difference in $T_2$ times between our study and real-world experiments.

As the spin density increases, the spins have a greater chance of coming into close proximity with each other. This implies that the local environment can exert a stronger influence on the Hamiltonian, particularly with regard to the transverse anisotropy that determines the qubit gap. Consequently, higher homogeneity leads to a greater standard deviation in our normal distribution of random qubit gaps. It is conceivable that even at a higher density, such as x=0.01, the $T_2$ time could remain similar to that at x=0.001.

Our results suggest the standard deviation of 0.007\% of E for x=0.001 and 0.3\% of E for x=0.01 in our normal distribution to bring the $T_2$ time closer to the experimental data. Additionally,At extreme density, $x=0.1$, 1.6\% of E can reproduce the experimental $T_2$ time of around 0.6$\mu s$. The last estimation of the standard deviation fall within a reasonable range, given that a $B^4_4$ term of Steven's operator has been estimated to have a full withd at half maximum (FWHM) of 1.6\% of $B^4_4$, which corresponds to a standard deviation of 0.68\% of $B^4_4$ at $x=0.1$ density of the Holmum crystal \cite{Shiddiq2016}.

We aim to investigate an additional aspect of an electron spin bath, specifically its spatial uniformity. Apart from the examination of the total number of spins, our placement of spins has been entirely random without any constraints. In our Monte Carlo simulations, this results in some spin configurations displaying a tendency for clustering, while others do not. To gain insights into the impact of spatial uniformity, we have employed a cubic cell, partitioned into seven segments, as illustrated in Figure \ref{fig:squarecell}. This partitioning ensures that no spins are positioned in close proximity to the central spin, promoting a relatively uniform spatial distribution within the cell.
\begin{figure}[ht]
    \centering
    \includegraphics[width=0.8\columnwidth]{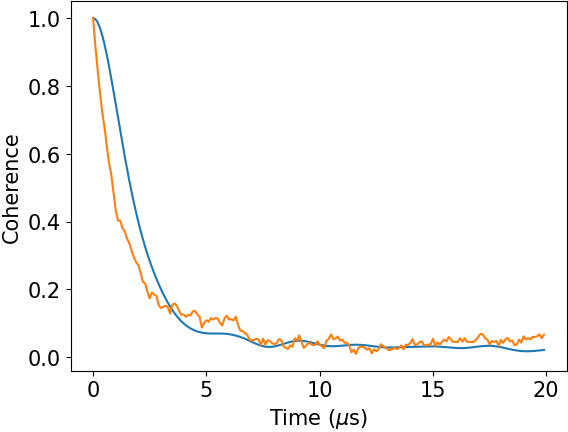}
    \caption{ The coherence signal in orange is generated from a set of completely random configurations. The blue coherence curve is a result from relatively uniform configurations. The latter $T_2$ time is about 53\% longer than the former.  }
    \label{fig:uniformity}
\end{figure}
To provide a basis for comparison, we have also generated seven spins randomly within a spherical box containing 7000 sites. Our results reveal a $T_2$ time of 1.45 $\mu s$ for the completely random configurations, while the uniform configurations exhibit a significantly longer $T_2$ time of 2.22 $\mu s$. This constitutes a 53\% increase in the $T_2$ time for the uniform configurations when compared to the completely random ones.  As depicted in Figure \ref{fig:uniformity}, we observed that the $T_2$ time of artificially arranged random configurations is longer when contrasted with completely random configurations. Our hypothesis is that spins with a relatively uniform distribution tend to exhibit a higher prevalence of low-frequency components in the signal. Consequently, this leads to longer coherence times when subjected to averaging. This finding could represent another important factor to consider when engineering multi-qubit devices.

\subsection{Cluster Correlation Expansion Methods Results}
In our previous observations, when two qubits possess nearly identical qubit gaps or gaps of comparable magnitude with respect to their interaction strength, the resulting signal exhibits a cosine-like behavior, and as long as the difference in gap values is smaller than the interaction strength, the signal can take on negative values.

A challenge arises when we compute the coherence function using the Cluster Coherent Excitations (CCE) technique. In this method, the coherence functions of subclusters are placed in the denominator, as depicted in equation \ref{eq:clusterexpansion}. This has an important implication: if a cluster, denoted as $C$, contains a subcluster, denoted as $C'$, and the coherence signal from $C'$ crosses zero, the quantity $\tilde{L}_C$ can experience a divergence as $L_{C'}$ approaches zero at the crossing time. This complication poses difficulties in the investigation of electron baths, as these divergences can distort the overall signal.

When we have a sufficiently large number of spins in the system, it becomes challenging to trace which clusters are responsible for generating these substantial peaks, and it is unclear how to rectify this issue. As a workaround, we employ a strategy where we generate random configurations, similar to our previous approach, and calculate the signals. However, if a result exhibits a substantial divergence or returns a 'NaN' (not a number) value, we discard that result and proceed to the next random configuration. We repeat this process until we accumulate a sufficient number of successful calculations for averaging.

To achieve a set of 200 successful calculations, we often require anywhere from a few thousand to several tens of thousands of random configurations. In the course of conducting a CCE simulation, the standard procedure involves determining the maximum CCE order and fine-tuning parameters such as "r-bath" and "r-dipole" through convergence tests. Here, "r-bath" represents a cutoff radius for bath spins that are considered in interactions with the central spin, and "r-dipole" represents a cutoff radius for forming clusters. Given that electron spin-spin interactions are significantly stronger than electron-nuclear spin-spin interactions, we typically need to use large values for "r-bath" and "r-dipole." Consequently, the number of clusters increases nearly exponentially as we raise these parameters and the maximum order.

We acknowledge that the parameters used in our CCE method results may not be fully converged due to the aforementioned challenges. Nevertheless, it is worth noting that the qualitative trends observed in the CCE results align with those obtained from exact diagonalization when dealing with a smaller number of total spins.

 \begin{figure}[ht]
    \centering
    \includegraphics[width=\columnwidth]{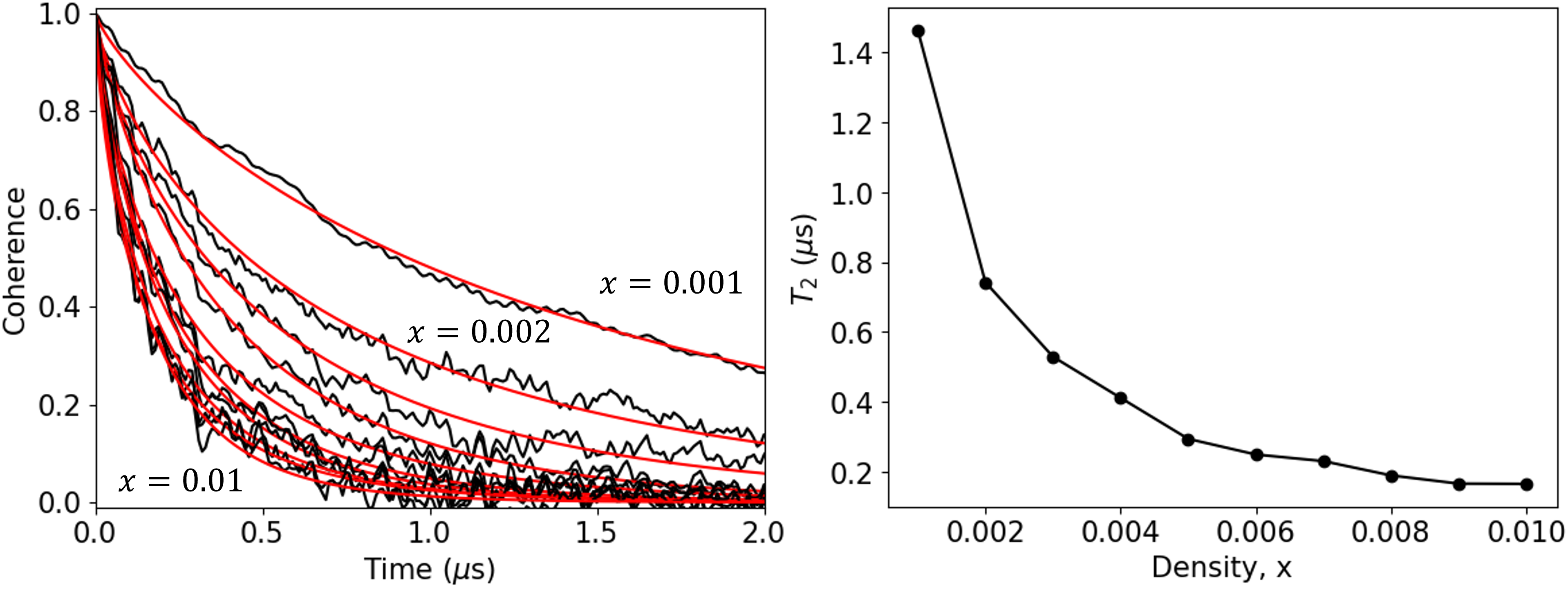}
    \caption{Left: coherence signals with different density of spin bath from CCE technique. The actual data are in black and the fitted curves in red. Right: $T_2$ time as a function of the density, $x$.}
    \label{fig:density_cce}
\end{figure}
\begin{figure}[ht]
    \centering
    \includegraphics[width=\columnwidth]{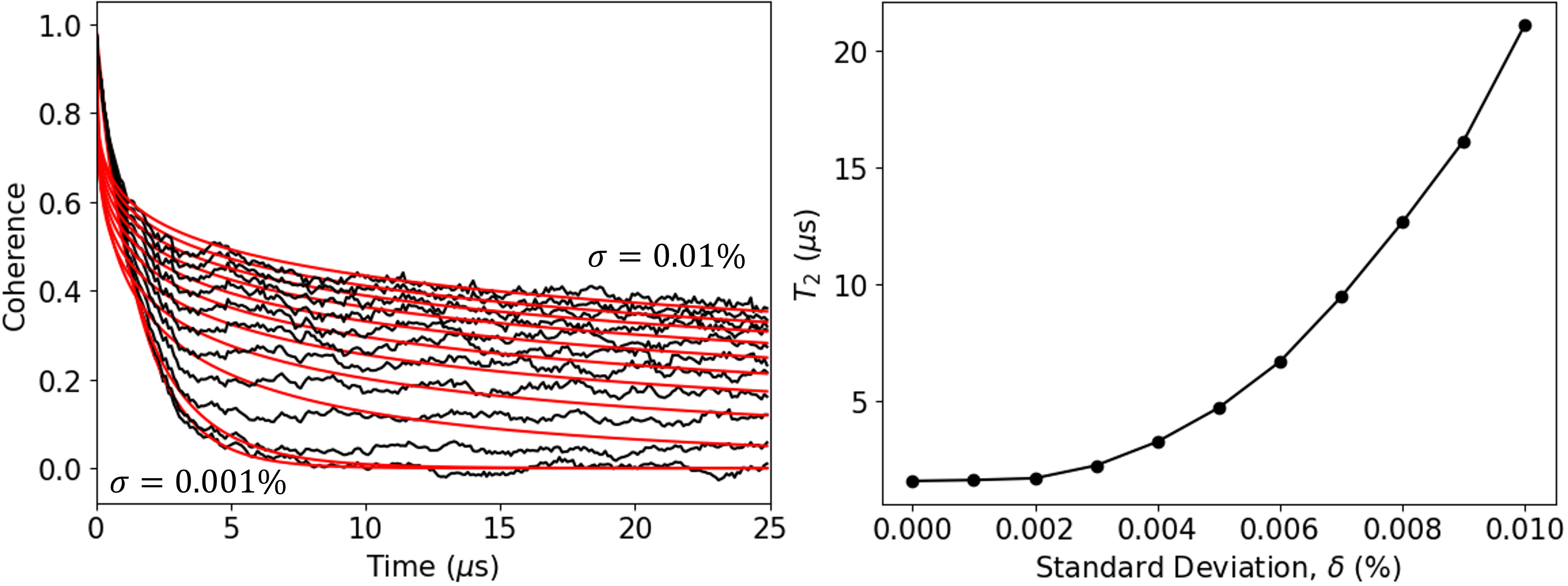}
    \caption{Left: coherence signals with different standard deviations of normal distributions of random qubit gaps from CCE technique. Right: $T_2$ time as a function of the standard deviation as a fraction of the original $E$.}
    \label{fig:off-resonance_cce}
\end{figure}

As depicted in Figures \ref{fig:density_cce} and \ref{fig:off-resonance_cce}, the findings generated through the CCE technique exhibit a congruent trend with those attained via exact diagonalization. In the context of the parameterization employed for these simulations, we have opted for a maximum CCE order of 3, an "r-bath" distance of 136.63 \AA, and an "r-dipole" separation of 40 \AA. These specific parameter selections have been made due to their close alignment with the coherence outcomes produced by exact diagonalization. However, it is important to note that modifying these parameters may yield different results.

\subsection{Conclusion}
In this chapter, our primary goal was to explore the impact of the electron spin bath on spin decoherence. To gain a clear perspective on this effect, we deliberately omitted the most substantial source of decoherence, the nuclear spin bath, by conducting our simulations precisely at the clock transition. For a more realistic approach, we derived structural and energetic parameters from the real-world material \ce{HoW10}.

Commencing our exploration with a focus on two-spin interactions, we unveiled fundamental insights into the interplay between two qubits. Our observations revealed that the coherence signal took the form of a cosine function, with its frequency mirroring the strength of interaction. Notably, we also identified that the signal's amplitude could diminish when qubits had mismatched qubit gaps, accompanied by corresponding frequency shifts.

To assess the impact of the electron spin bath's density, we initially employed exact diagonalization. This involved creating a larger cell and systematically varying its size while keeping the number of spins constant within the system. Our findings unveiled a compelling trend: denser spin baths were associated with more rapid decoherence of the central spin.

Furthermore, we delved into simulating the effect of inhomogeneity within the qubit gaps, inspired by our observations with two qubits. Our analysis indicated that systems displaying greater inhomogeneity in the qubit gap were better equipped to retain coherence for more extended periods.

Another dimension of our investigation revolved around the spatial uniformity of spins within the bath. We explored this aspect by comparing results obtained from completely random spin configurations with those from configurations that, although still random, exhibited a relatively more uniform distribution of spins. Our findings highlighted that configurations with enhanced uniformity were linked to extended coherence times.

Lastly, we presented outcomes from the CCE technique, featuring specific parameter choices. It is noteworthy that future studies may delve more deeply into the nuances of parameter selection, offering further insights into these intricate dynamics.

\begin{acknowledgments}
The authors are grateful for useful conversations with Xiao-Guang Zhang and Steve Hill. This work is supported by the U.S. Department of Energy, Office of Science, Basic Energy Sciences under Award No.{} DE-SC0022089 and Award No.{} DE-SC0019330. The calculations were performed using the utilities of the National Energy Research Scientific Computing Center and the University of Florida Research Computing.
\end{acknowledgments}

%\clearpage
\bibliographystyle{apsrev4-2}
\bibliography{main.bib}

\end{document}